# The Linear $U_S$-$u_P$ Relation in Shock-Wave Physics

A Kerley Technical Services Research Report

Gerald I. Kerley, Consultant

March 2006

A narrated PPT tutorial on this subject can be found at
http://kerleytechnical.com/tutorials.htm

# The Linear $U_S$-$u_P$ Relation in Shock-Wave Physics

A Kerley Technical Services Research Report

Gerald I. Kerley, Consultant

March 2006

## ABSTRACT

A linear relation between shock velocity $U_S$ and particle velocity $u_P$ is often regarded as the "typical" or "standard" material response in the shock-wave literature. It has even been proposed that this linearity follows from some kind of universal equation of state (EOS) principle. This report presents a theoretical analysis of this issue and a survey of the Hugoniot data for all the elements. It demonstrates that linearity follows from the fact that $U_S$-$u_P$ plots are rather insensitive to material properties, not from any universal EOS. The effects of pressure and material properties on the shock response are more easily seen and analyzed by plotting $U_F = U_S - u_P$ as a function of $u_P$. The data survey shows that linear behavior is only observed in 20% of all the elements and is not at all universal.

# CONTENTS

# FIGURES

# SYMBOLS AND UNITS

| | |
|---|---|
| $\rho$ | density [$g/cm^3$] |
| $P$ | pressure [$GPa$] |
| $E$ | specific internal energy [$MJ/kg$] |
| $U_S$ | shock velocity (in laboratory frame) [$km/s$] |
| $u_P$ | particle velocity (in laboratory frame) [$km/s$] |
| $U_F$ | $U_S - u_P$, shock velocity in moving frame [$km/s$] |
| $C_0, S$ | coefficients in linear relation $U_S = C_0 + Su_P$ |
| $\varepsilon_k$ | coefficients in relation $U_F = C_0 + \varepsilon_1 u_P + \varepsilon_2 u_P^2 + \dots$ |
| $\rho_0$ | density of solid at zero pressure and temperature [$g/cm^3$] |
| $\mu$ | $1 - \rho_0 / \rho$ strain [$unitless$] |
| $\nu$ | $\rho / \rho_0 - 1$ [$unitless$] |

# PREFACE

The shock-wave literature abounds with tributes to the linear $U_S$-$u_P$ relation. Consider the following example:

> "In the field of shock compression science it has been found that for an overwhelming number of solid and liquid materials, the velocity $U_S$ of a planar, compressive shock traveling into unstrained material at rest is linearly related to the material velocity behind the shock $u_P$."

The above quotation appears in a paper that listed me as a coauthor—without my knowledge or approval. (I have corrected errors that appeared in the original.)

In fact, I disagree with the above remark and other claims that most materials are described by linear $U_S$-$u_P$ relations. However, I have been too busy with other matters to develop arguments in support of my position.

That situation has changed, now that I am giving myself time to work and reflect on matters that really interest me. This report is the result of my reflections about linear $U_S$-$u_P$ relations.

Having worked in this field for 37 years, I have no illusions that this report will receive a warm welcome from the shock-wave community. However, I do maintain that anyone who actually takes the time to read it—with an open mind—will find my arguments sound and worthy of consideration.

Gerald I. Kerley
Appomattox, VA
March, 2006

# 1. INTRODUCTION

A linear relationship between the shock and particle velocities is frequently used to represent experimental Hugoniot data in a variety of shock-wave applications. This relation is usually written in the form

$$U_S = C_0 + S u_P, \tag{1}$$

where the parameter $C_0$ is roughly equal to the zero-pressure bulk sound speed[1], and the parameter $S$ typically ranges from 1.0 to 1.6.

Equation (1) was first introduced in the classic 1958 paper of Rice, McQueen, and Walsh [1][2], and used to fit data for 23 metals. However, Hugoniot data tabulations, such as the Los Alamos shock wave compendium [3], offer linear fits for all kinds of materials, including metals and alloys, minerals, plastics, and even unreacted explosives. The linear relationship is known to break down in some cases—materials with phase transitions, porosity, large strength effects, or molecular bonding [2]. But it is often used even in those cases, at least for representing the data over part of the range.

Today, nearly fifty years after its introduction, the "linear $U_S$-$u_P$ curve" has become so widely used and accepted that it is sometimes regarded as a kind of "law" of shock wave physics. A reported deviation from linearity is often treated with skepticism, requiring some kind of "mechanism" to explain its existence. It has even been proposed that the linear relationship indicates some kind of universality in the equation of state (EOS). Papers attempting to "prove" the linear relationship, using various EOS models, appear from time to time.

In this report, I will demonstrate that a linear $U_S$-$u_P$ relation does *not* follow from any universal EOS. On the contrary, it follows from the fact that a plot of $U_S$ vs. $u_P$ is actually *insensitive* to variations in material properties. This fact becomes more evident when the shock- and particle-velocity data are presented in a form that is more sensitive to the EOS. In that case, the linear relationship turns out to be much less prevalent than previously thought.

It is clear that the existence of a shock wave implies that $U_S > u_P$. Equation (1) shows that this condition implies $S \geq 1$; if $S < 1$, the unphysical condition $U_S < u_P$ occurs for $u_P > C_0/(1-S)$. Of course, the condition $S < 1$ is allowed over a *limited* particle velocity range; some examples of that situation will be discussed in this report. But the occurrence of that condition shows that the Hugoni-ot is *not* linear over its entire range; it *must* show upward curvature at high pressures, although that curvature may occur outside the range of the available data.

---

1. $C_0$ is usually larger than the zero pressure sound speed when material strength effects are important. It is usually less in porous materials.

Hence $S \geq 1$ is a necessary condition for the existence of a linear $U_S$-$u_P$ relationship. This result follows only from the definition of a shock wave, not from any EOS principle. One might suspect that variations in material properties can be seen more easily by comparing $S - 1$ instead of $S$. However, that approach does not shed light on the issue of linearity.

This report offers another way to examine the effects of EOS variations on shock wave data. Most experiments measure $U_S$, the shock velocity in the laboratory frame of reference. A different, but equally valid, measure of the shock velocity is,

$$U_F = U_S - u_P, \text{ which is equivalent to } U_S = U_F + u_P. \quad (2)$$

$U_F$ is just the shock velocity in a frame of reference moving with the material behind the shock, i.e., with velocity $u_P$.

In order to create some familiarity with this quantity, and to lay the foundation for the rest of the report, Sec. 2 discusses the "reverse ballistic" point of view, in which $U_F$ appears in a natural way. It also discusses the advantages of plotting $U_F$ vs. $u_P$, instead of $U_S$ vs. $u_P$, to examine the dependence of shock data on material properties.

Section 3 examines expressions for the Hugoniot in the pressure-density plane. It shows that a realistic EOS, for any type of material, leads to the condition $S > 1$. It also demonstrates that deviations from a linear $U_S$-$u_P$ relation are especially likely when $S > 4/3$. Additional discussion is given in Appendix A.

Section 4 presents the results of a survey of Hugoniot data for the elements. This survey shows that *only 20%* of the cases for which data exist show generally linear behavior. Examples of non-linear behavior are presented for Al, Cu, Pb, W, Ta, Sc, Y, La, Ca, and Au.

Conclusions are summarized in Sec. 5.

Throughout this report, I will be considering the simple case of a single, one-dimensional, steady, planar shock wave. To avoid repetition, the terms "shock," "shock wave," "Hugoniot," etc., will always refer to this case.

## 2. REVERSE BALLISTIC VIEW

Figure 1 compares two equivalent ways of describing shock-wave propagation. In the conventional picture (Fig. 1a), a rigid (incompressible) piston with velocity $u_P$ generates a shock with velocity $U_S$ in a material initially at rest. In this picture, $u_P$ is both the piston velocity and the particle velocity of the material behind the shock front. $U_S$ is the shock velocity in the laboratory frame.

In the reverse ballistic picture (Fig. 1b), the unshocked material, with an initial velocity $u_P$, impacts a rigid, stationary anvil, generating a shock with velocity $U_F$. Hence $U_F$ is just as valid a measure of the shock velocity as $U_S$ and contains all the same information about the material properties. However, we will see that it is more sensitive to variations in those properties than $U_S$.

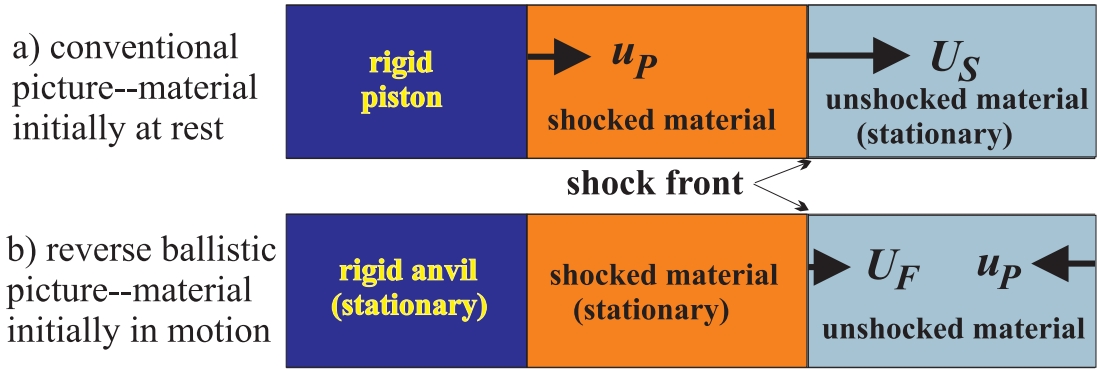

**Fig. 1. Two equivalent depictions of one-dimensional shock propagation. In the "conventional" view (a), the unshocked material is intially at rest; a rigid piston with velocity $u_P$ generates a shock wave with velocity $U_S$ into the material. In the "reverse ballistic" view (b), the unshocked material is initially in motion with velocity $u_P$; it impacts a rigid anvil, generating a shock with velocity $U_F = U_S$-$u_P$**

$U_F$ can also be regarded as the velocity at which the shock front "outruns" the piston. Since $U_S > u_P$ and $U_F > 0$ for all materials, independent of any EOS effects, a plot of $U_F$ vs. $u_P$ contains the same information as a plot of $U_S$ vs. $u_P$.

It is also clear that $U_F$ and $U_S$ have the same intercept $C_0$ as $u_P \rightarrow 0$. In the absence of material strength and porosity effects, $C_0$ is just the zero pressure bulk sound speed. Let us write the dependence of $U_F$ on $u_P$ as

$$U_F = C_0 + \varepsilon_1 u_P + \varepsilon_2 u_P^2 + \ldots = C_0 + \xi(u_P). \tag{3}$$

If the constants $\varepsilon_k$ are small, i.e, if $U_F$ is essentially constant over the range of measurements, plotting $U_F$ vs. $u_P$ immediately reveals that the data do not contain any EOS information except the sound speed. By contrast, a plot of $U_S$ vs. $u_P$ seems to indicate a "normal" linear material, even though it does not offer any more information; linearity is guaranteed because, apart for a constant shift, one is simply plotting $u_P$ vs. itself.

It is easy to see that $U_F$ is always a much weaker function of $u_P$ than $U_S$, since the Hugoniot slope is given by $\tilde{S} \approx 1 + \varepsilon_1$, $S$ typically ranging from 1.0 to 1.6.

The above observation leads to the main thesis of this report: When $U_F$ is a weak function of $u_P$, as it is for most materials, a $U_S$-$u_P$ plot is expected to be generally linear because it is, in part, just a plot $u_P$ vs. itself. Consequently, a $U_S$-$u_P$ plot obscures deviations from linearity that are more easily seen in a $U_F$-$u_P$ plot.

It is well known that all materials look rather similar when the Hugoniot data are plotted in the $P$-$u_P$ plane. (Most theorists know that even a poor EOS will look credible in that type of plot.) It is easy to explain this phenomenon by noting that the Hugoniot pressure is given by $P_H = \rho_0 U_S u_P$; plotting $P_H$ vs. $u_P$ further dilutes the material information in a $U_S$-$u_P$ plot by multiplying the dependent variable by the independent variable.

$U_S$-$u_P$ and $P$-$u_P$ plots do have legitimate uses, of course. $U_S$-$u_P$ plots are useful for understanding the effects of multiple waves, generated by elastic precursors and phase transitions, on shock propagation. $P$-$u_P$ plots are indispensable for understanding phenomena associated with shock propagation across material interfaces, especially in the analysis of impedance-match experiments. This report is not intended to dismiss such plots as useless, only to expose their limitations in revealing the material property information contained in shock data.

## 3. PRESSURE-DENSITY FORMULAS

For the linear $U_S$-$u_P$ relation, Eq. (1), the pressure-density formula is

$$P_H = \rho_0 C_0^2 \mu / (1 - S\mu)^2, \text{ where } \mu = 1 - \rho_0 / \rho. \tag{4}$$

This expression is valid only if $\mu < 1/S$, or

$$\rho < \rho_A = \rho_0 S / (S - 1), \tag{5}$$

where $\rho_A$ is the "Hugoniot asymptote."

The fact that $S$ appears in Eq. (4) (as opposed to $\varepsilon_1 = S - 1$, for example) may suggest that $U_S$, not $U_F$, is the more fundamental quantity. In this section I will show that this impression is misleading.

First let us express the pressure in terms of the quantities $\varepsilon_k$ in Eq. (3). The result for the linear case is

$$P_H = \rho_0 C_0^2 (\nu + \nu^2) / (1 - \varepsilon_1 \nu)^2, \text{ where } \nu = \rho / (\rho_0 - 1).^1 \tag{6}$$

In Sec. 1, I noted that the condition $S < 1$ ($\varepsilon_1 < 0$) leads to the unphysical result $U_S < u_P$ at high densities. Examination of Eq. (6) shows it also gives the unphysical result $P_H \to \rho_0 C_0^2 / \varepsilon_1^2 = $ constant as $\rho \to \infty$. It is possible for a material to have a slope $S < 1$ over *part* of the range of the data. Examples of this situation will be given in Sec. 4. However, the very existence of such a condition demonstrates that the Hugoniot must have upward curvature at high densities.

The condition $S = 1$ ($\varepsilon_1 = 0$) gives a quadratic dependence of the Hugoniot pressure on density. This result is not unphysical, but it is unrealistic; no reasonable theoretical EOS model would give such a result.[2] Here again, it is possible to have $S = 1$ over part of the range, but it must curve upward at high densities.

The above results show that $S > 1$ ($\varepsilon_1 > 0$) for any reasonable Hugoniot EOS, for *any* material. It follows that all of the material-dependent information is contained in the function $U_F$, as claimed in the previous discussions.

---

1. For completeness, the result for the quadratic case is
$$P_H = \rho_0 C_0^2 (\nu + \nu^2) / \left[ (1 - \varepsilon_1 \nu)/2 + \sqrt{(1 - \varepsilon_1 \nu)^2/4 - \varepsilon_2 \nu^2 / C_0} \right]^2.$$
For the purposes of this report, there is little insight to be obtained from considering higher-order expressions.

2. See Table 5 of Ref. [4] for a partial listing of formulas often used in EOS modeling.

Further insight can be obtained from consideration of the Hugoniot asymptote. Appendix A derives the well known ideal-gas value, $\rho_A = 4\rho_0$, which is the limit for all materials at sufficiently high density. This condition corresponds to $S = 4/3$, from Eq. (5).

The approach to the asymptote can be complicated. Appendix A shows Hugoniots for five materials that cross the ideal gas value and approach it from the high-density side. This result, which is typical for most materials, results from deviations from the ideal gas EOS because of incomplete ionization.

Examination of Eq. (5) shows that the condition $\rho/\rho_0 > 4$ implies a Hugoniot slope $S < 4/3$. It follows that any material having a slope $S > 4/3$ is virtually certain to exhibit downward curvature at high shock pressures. Of course, this observation does not rule out the possibility that materials having a slope $S < 4/3$ will also exhibit curvature.

The findings in this section demonstate that the linear $U_S$-$u_P$ relation is not at all universal. Deviations from linearity are actually expected in most materials.

# 4. EXAMPLES

A comprehensive survey of the shock wave literature would be far beyond the scope of this report. However, I have reviewed the data for the 92 elements up through U. The unclassified literature [3][5][6] offers data for 75 elements—59 metals (including the lanthanides and actinides) and 16 non-metals (including semi-conductors and gaseous elements).[1] The following observations were made.

- None of the 16 non-metals[2] can be represented by linear $U_S$-$u_P$ plots. Deviations from linearity are due to phase transitions, the effects of molecular bonding, and, in the case of Xe [7], to electronic rearrangements.

- None of the 14 lanthanides (Ce to Lu) can be represented by linear $U_S$-$u_P$ plots. Deviations from linearity are due to phase transitions and/or electronic rearrangements.

- Eight other metals[3] cannot be represented by linear $U_S$-$u_P$ plots because of phase transitions with significant volume changes.

- 18 other metals[4] show definite deviations from linearity, especially visible in $U_F$-$u_P$ plots.

- 15 metals[5] show generally linear behavior, at least within the range and scatter of existing data.

- The data for four other metals[6] are too sparse or exhibit too much scatter to make a reliable judgment about linearity.

In summary, only 15 out of 75 elements—20% of the cases for which data exist—show generally linear behavior. And most of these materials would be expected to show non-linear behavior if the range of data were increased. A linear relation between $U_S$ and $u_P$ is clearly *not* a universal property of shock-wave data.

Figure 2 compares plots of the two shock velocities, $U_S$ and $U_F$, as functions of particle velocity $u_P$, for aluminum—the most widely studied material in the shock-wave literature. (The circles are experimental data for high-purity Al alloys (1100, 3300, and 6066) [3][5][8][9].)

As expected, $U_S$ (Fig. 2a) shows a much stronger dependence on particle velocity than $U_F$ (Fig. 2b). In fact, $U_F$ accounts for only 25% of the dependence of $U_S$ on $u_P$; the remaining 75% is just due to the contribution from $u_P$ itself.

---

1. No shock wave data could be found for the following elements: F, Cl, Br, Ne, Rn, As, Te, At, Mn, Tc, Ru, Os, Po, Fr, Ra, Ac, Pa.
2. H, He, C, N, O, Si, P, S, Ar, Ge, As, Se, Kr, I, Xe. Existing data for a few of these materials are sparse, but a judgement about linearity can be made from theoretical considerations.
3. Bi, Cs, Fe, Hf, Rb, Sn, Ti, Zr.
4. Al, Au, Ba, Be, Ca, Cu, La, Pb, Pd, Sc, Sr, Ta, Tl, U, V, W, Y, Zn.
5. Ag, Cd, Co, Cr, In, Ir, K, Li, Mg, Mo, Na, Nb, Ni, Pt, Rh.
6. Ga, Hg, Re, Th.

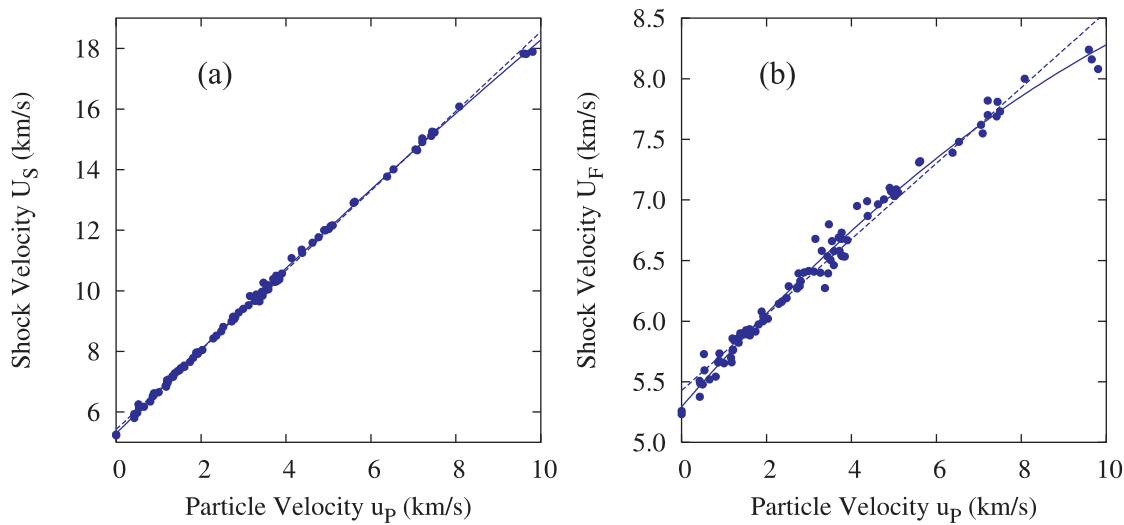

**Fig. 2.** Hugoniot data for aluminum. Circles are experimental data [3][5][8][9]. Solid lines show quadratic fits, dashed lines show linear fits..

Figure 2 also shows quadratic fits to the data (solid lines), and linear fits (dashed lines).[1] Examination of Fig. 2a gives the general impression that the linear fit is an adequate representation of the $U_S$-$u_P$ behavior, the quadratic fit offering only marginal improvement. But the $U_F$-$u_P$ plot, Fig. 2b, clearly shows the superiority of the quadratic fit, with downward curvature at high pressures.

Figure 3 compares the $U_F$-$u_P$ data with the Hugoniot computed from a theoretical EOS [11]. The theoretical model agrees with the presence of downward curvature but indicates that a simple quadratic fit is oversimplified. Careful examination of Fig. 3 shows a slight increase in $U_F$ for $u_P$ in the range 4-4.5, where the Hugoniot crosses the melting curve. At higher values of $u_P$, $U_F$ again shows generally linear behavior but with a smaller slope than in the solid region. These details are more difficult to detect in the $U_S$-$u_P$ plot.

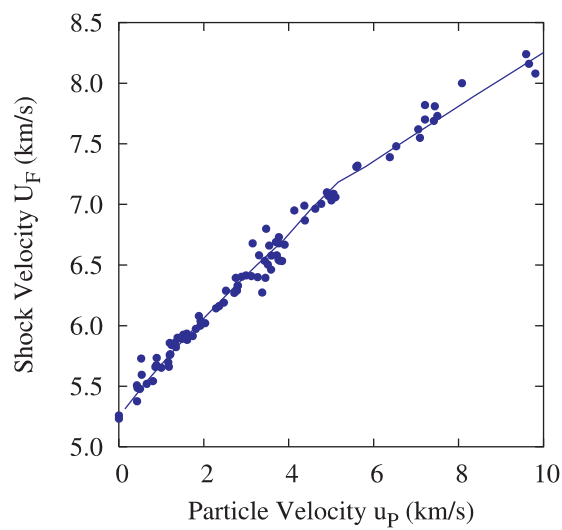

**Fig. 3.** Comparison of theory with experiment for Al. Experimental data are the same as in Fig. 2. The solid line was calculated from a theoretical EOS [11] .

It should be noted that all of the data in Figs. 2 and 3, for $u_P > 6$, were taken from Ref. [9]. Before that paper was published, in 2003, data in the high-pressure region were sparse and exhibited significant scatter. The theoretical EOS, published in 1987, agrees with the existence of nearly linear behavior in the data for $u_P < 6$, even in the $U_F$-$u_P$ plot. Hence aluminum illustrates the error of using a linear extrapolation to predict the EOS outside the range of existing data.

---

1. The fits discussed in this report were made using the gnuplot plotting utility [10].

Downward curvature can also be seen in the $U_F$-$u_P$ plots for copper and lead, shown in Figure 4. Blue circles show the experimental data for Cu, red squares show the data for Pb [3][5][8]. The solid lines in Fig. 4a are quadratic fits to the data and the dashed lines are linear fits. The quadratic fits are clearly superior to the linear ones, especially in the case of Pb.[1]

The case for curvature is strengthened when the experimental data are compared with Hugoniots computed from the theoretical EOS [12], shown by the solid lines in Fig. 4b. As in Al, the simple quadratic fits give an oversimplified description. In Cu, melting leads to a slight increase in $U_F$ for $u_P$ in the range 2.7-3.3, followed by a decrease in slope in the liquid region at higher pressures. In Pb, melting causes an increase in $U_F$ for $u_P$ in the range 1.1-1.4, although it is difficult to see in Fig. 4b. The curve for Pb shows not only a decrease in slope at melting, but also pronounced downward curvature with increasing pressure.

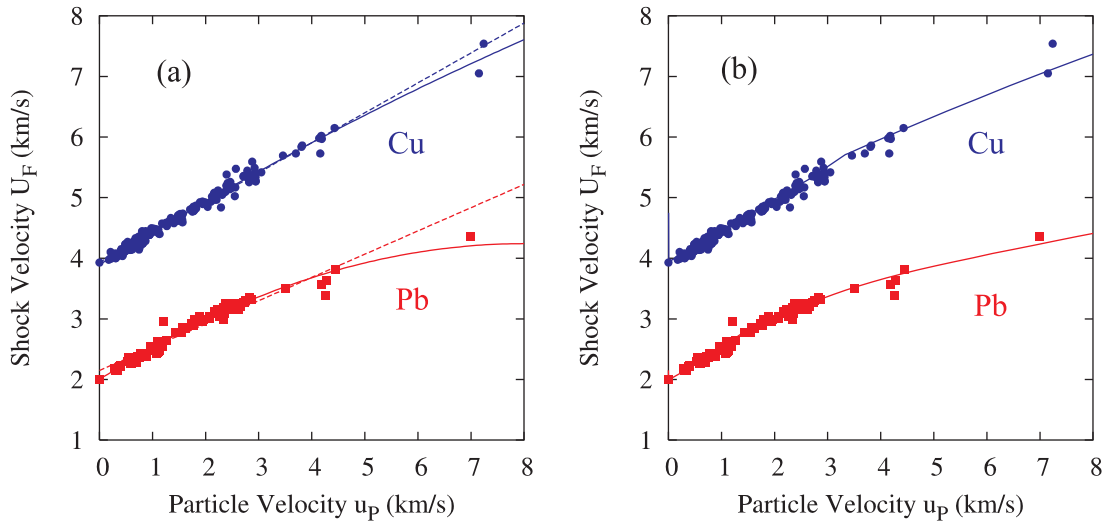

**Fig. 4. Hugoniot data for copper and lead. Experimental data [3][5][8]: blue circles—Cu; red squares—Pb. In (a), solid lines show quadratic fits, dashed lines show linear fits. The solid lines in (b) were calculated from theoretical EOS [12].**

---

1. For Cu, it could be argued that the experimental data for $u_P > 5$ are too sparse to warrant any firm conclusions. However, data at even higher pressures agree with the existence of curvature. In fact, the shock compilation of Trunin, et al. [6], recommends a quadratic fit to the $U_S$-$u_P$ curve for Cu.

It should not be assumed that deviations from linearity always result in *downward* curvature. The opposite effect is illustrated in Figs. 5 and 6.

Figure 5 shows $U_S$-$u_P$ and $U_F$-$u_P$ plots for tungsten and tantalum. Blue circles show the experimental data for W, red squares show the data for Ta [3][5][8][13][14]. For W, the solid lines were computed from using a theoretical EOS [15], the melting transition shown by a dashed line. Material strength was included in this calculation, and the elastic precursor is also shown. For Ta, the solid lines were generated by a cubic fit to the data because I have not yet generated a theoretical EOS for that material.

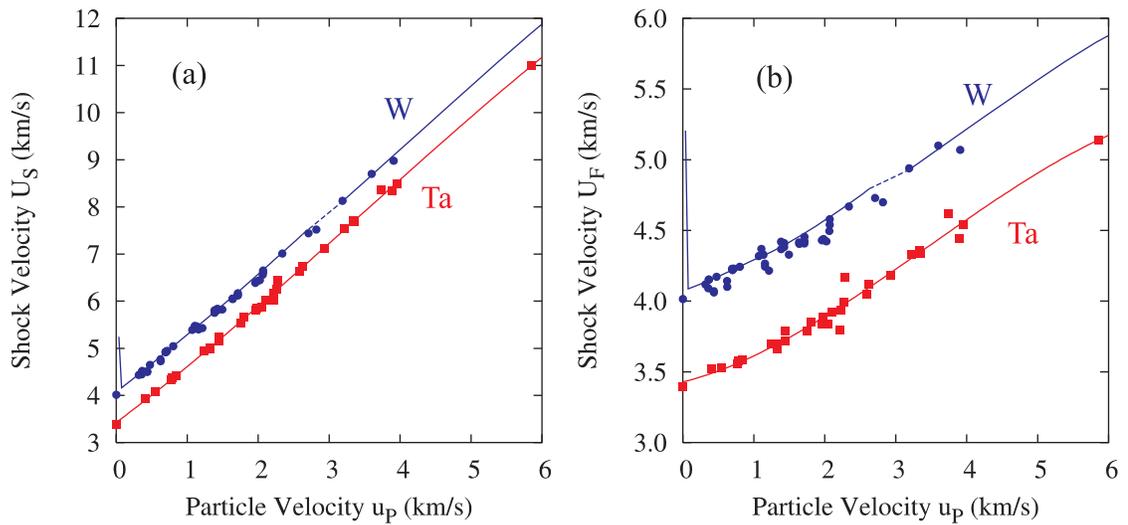

**Fig. 5. Hugoniot data for tungsten and tantalum. Experimental data [3][5][8][13][14]: blue circles—W; red squares—Ta. For W, the lines were calculated from a theoretical EOS [15], the melting transition shown by a dashed line. For Ta, the lines were computed from cubic fits.**

W and Ta are normally regarded as linear materials, and the $U_S$-$u_P$ plots (Fig. 5a) do give the general impression of linearity. However, $U_F$ accounts for only 20% of the dependence of $U_S$ on particle velocity in these two materials. 80% of the apparent linearity in the $U_S$-$u_P$ plots results from plotting $u_P$ vs. itself, which obscures the non-linear effects.

By contrast, the $U_F$-$u_P$ plots (Fig. 5b) show pronounced *upward* curvature, with a "rounding off" and the beginning of downward curvature at the highest pressures. This behavior is also captured quite well by the theoretical model for W and is evident in the cubic fit for Ta.

Figure 6 shows $U_S$-$u_P$ and $U_F$-$u_P$ plots for the Group IIIB metals, scandium, yttrium, and lanthanum. Blue circles show the experimental data for Sc, red squares show the data for Y, and green triangles show the data for La [3][5]. In this case, deviations from linearity can be seen even in the $U_S$-$u_P$ plots (Fig. 6a). However, the $U_F$-$u_P$ plots (Fig. 6b) are more interesting and reveal important details. For La, $U_F$ is nearly constant for $u_P < 1$, then shows sharp upward curvature at higher pressures. For Sc and Y, $U_F$ initially decreases with $u_P$ before curving upward at high pressures. This initial decrease in $U_F$ corresponds to the condition $S < 1$, which cannot persist to high pressures, where it would lead to unphysical behavior. Hence the upward curvature observed in the data is actually required by theoretical considerations.

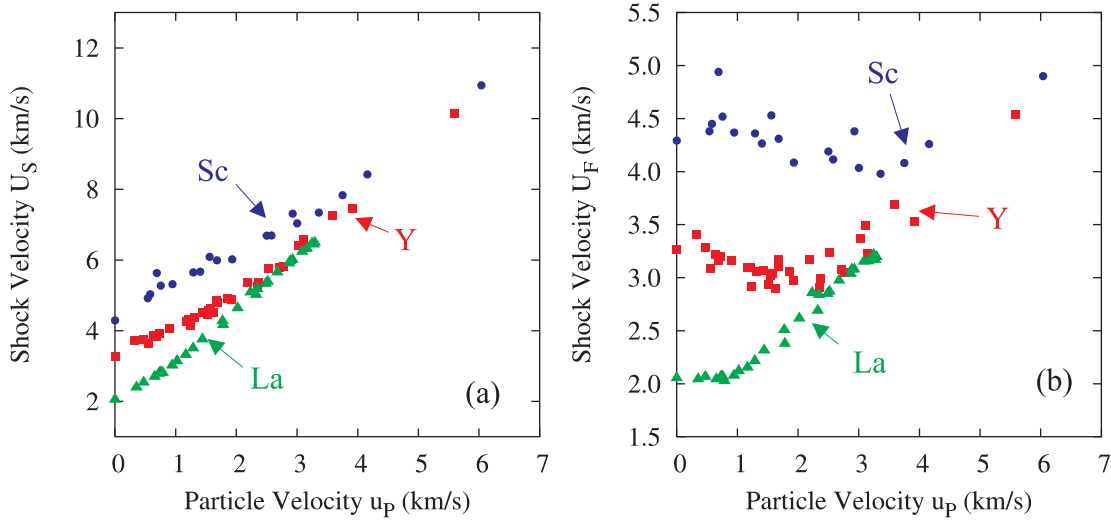

Fig. 6. Hugoniot data for scandium, yttrium, and lanthanum. Experimental data [3][5]: blue circles—Sc; red squares—Y; green triangles—La.

There is no evidence for pressure-induced phase transitions in either Sc or Y. La is believed to have a phase transition at low pressures, but the volume change is not expected to be significant because both phases have close-packed structures. Theoretical considerations indicate that the non-linear behavior is related to changes in the electronic structure, together with melting.

There is no reason to assume that the behavior seen in Fig. 6 is limited to the IIIB metals. Figure 7 shows a $U_F$-$u_P$ plot of the experimental data for calcium [3][5]. With the exception of a single data point, at $u_P = 7$, $U_F$ is seen to decrease with increasing particle velocity,[1] just as observed in Sc and Y. Indeed, the Los Alamos shock compendium represents the Ca data by a linear $U_S$-$u_P$ fit with $S = 0.94$ [3]. But, as we have now noted several times, the condition $S < 1$ corresponds to unphysical behavior, requiring the Hugoniot to curve upward at high pressures. The single point at $u_P = 7$ may be an indication of that curvature.

---

1. The data do not extrapolate to the bulk sound speed at $u_P = 0$, probably due to strength effects.

Figure 8 shows the last example of this report, a $U_F$-$u_P$ plot of the Hugoniot for gold. The experimental data [3][5], shown by circles, are surprisingly sparse and limited in range. In this case, a linear fit, shown by a dashed line, gives a reasonable representation of the data. A quadratic fit (not shown) differs only slightly from the linear one and shows no pronounced curvature. Based on the data alone, one might well conclude that Au is a linear material.

However, comparison with a theoretical EOS [15], shown by the solid line, shows that such a conclusion would be incorrect. The theoretical curve shows the same features seen in Al, Cu, and Pb (Figs. 2-4)—an increase in $U_F$ going through the melt transition, followed by a smaller slope and downward curvature in the liquid region. Once again, using the assumption of linearity to extrapolate the Hugoniot data to higher pressures would give erroneous results.

Admittedly, I have given considerable weight to the theoretical predictions in making the above arguments. I accept the fact that experimentalists might want to test these ideas. In fact, I encourage them to do so. New Hugoniot data for gold would be welcome.

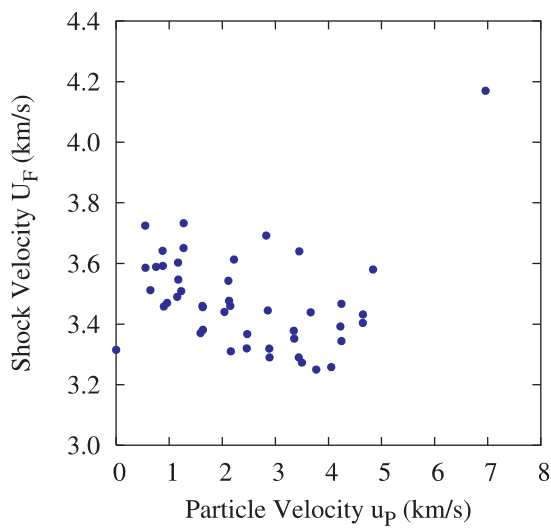

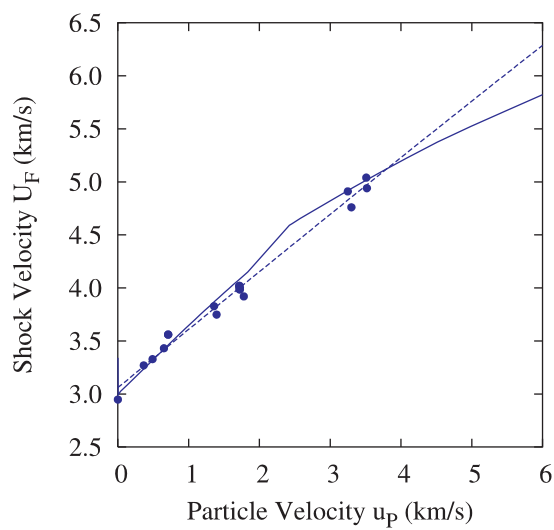

**Fig. 7.  Hugoniot data for calcium. Circles are experimental data [3][5].**

**Fig. 8.  Hugoniot data for gold. Circles are experimental data [3][5]. Solid line is a theoretical EOS [15]. Dashed line is a linear fit to the data.**

# 5. CONCLUSIONS

The principal conclusions of this report can be summarized as follows.

- $U_S$-$u_P$ plots often appear to be linear because they are, in large part, plots of $u_P$ vs. itself.

- Theoretical considerations show that deviations from linearity are required when $S < 1$ and are especially likely when $S > 4/3$.

- The effects of pressure and material properties on the shock response are more easily seen and analyzed by plotting $U_F = U_S - u_P$ vs. $u_P$.

- A survey of Hugoniot data shows that only 20% of the elements exhibit linear behavior.

- The survey shows examples of downward curvature, upward curvature, and more complicated deviations from non-linear behavior.

If one accepts these conclusions, what are the consequences? How should one proceed? I will finish this report with a few suggestions.

First, those who insist on using linear $U_S$-$u_P$ models should at least be aware of their limitations and not make overblown claims as to their validity. Knowledge of the facts could help people avoid certain common mistakes. One such mistake is to obtain a few data points on a material, fit it to a straight line, and assume the model has been "validated." Another mistake is to make a $U_S$-$u_P$ fit to data for a porous material and use it in a Mie-Grüneisen model; the correct approach is to fit the data for the TMD material for use with a pore-compaction model.

Second, the time has come to go beyond Mie-Grüneisen models, whether they are based on linear $U_S$-$u_P$ fits or more complicated expressions. In order to create good equations of state, one must face the need to use EOS modeling codes [18] and the EOS tables they generate.

Unfortunately, one cannot assume that an EOS table is reliable simply because it can be found in some database. Existing databases contain many tables that were generated with simplistic models and are inadequate for use in applications that require accurate descriptions of material behavior.

In order to generate a good EOS, one must be able to model a wide range of physical and chemical phenomena that affect material behavior—polymorphic phase transitions, melting and vaporization, molecular degrees of freedom, molecular dissociation and other chemical reactions, and electronic excitation and ionization [19]. The tools for treating these phenomena are already available to those who will take the time to learn and use them [18].

# Appendix A

# The Ideal Gas Asymptote

At sufficiently high temperatures, where all molecules are dissociated and all atoms ionized, all materials have the ideal gas EOS,

$$P(\rho, E) = (2/3)\rho E.$$  (A.1)

The energy conservation law for a shock wave is

$$E = E_0 + (1/2)(P + P_0)(1/\rho_0 - 1/\rho),$$  (A.2)

where $E_0$ and $P_0$ are the initial energy and pressure. Combining these two equations, and taking $E_0 = P_0 = 0$, one readily finds

$$\rho/\rho_0 = 4.$$  (A.3)

This result does not give any information about the *approach* to the asymptote. Figure 9 shows theoretical Hugoniots for five materials [12][15]-[17], in the $P$-$\rho$ plane, at temperatures up to $1.0\times10^8$ K. In all five cases, the Hugoniots cross the ideal gas asymptote and approach it from the high-density side. As noted in Sec. 3, the condition $\rho/\rho_0 > 4$ implies a slope $S < 4/3$ in the $U_S$-$u_P$ curve.

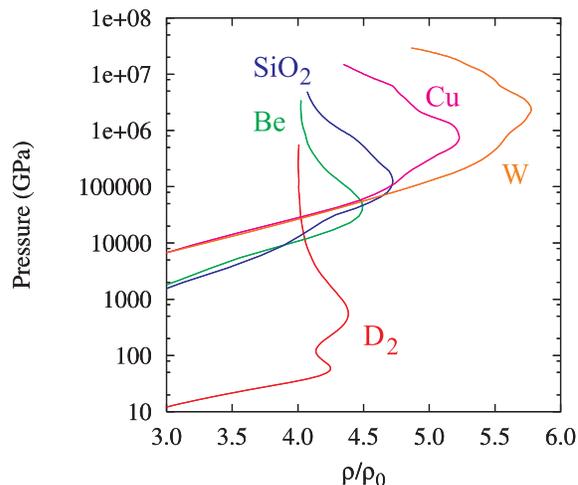

**Fig. 9. Approach to Hugoniot asymptote for five materials. The highest temperature is $1.0\times10^8$K in all cases.**

Only $D_2$ and Be, which have the lowest atomic numbers, actually reach the asymptote in Fig. 9, while $SiO_2$ (quartz) comes quite close. The fact that Cu and W have not yet reached the asymptote at the highest temperature, $1.0\times10^8$K, indicates that they are not yet fully ionized.

The first density maximum in the $D_2$ Hugoniot arises from molecular dissociation, the second from ionization. The $SiO_2$, Cu, and W Hugoniots have structure that arises from multiple stages of ionization.